\documentclass[aps,prb,twocolumn,superscriptaddress,floatfix]{revtex4-2}
\usepackage[utf8]{inputenc}
\usepackage{natbib}
\usepackage[colorlinks=true,linkcolor=blue,citecolor=blue,urlcolor=blue]{hyperref}
\usepackage{graphicx}
\usepackage{latexsym}
\usepackage{amssymb}
\usepackage{amsmath}
\usepackage{amsfonts}
\usepackage{amsthm}
\usepackage{bm}
\usepackage{floatrow}
\usepackage[caption=false]{subfig}
\usepackage{bbm}
\usepackage{enumitem}
\usepackage{hyperref}
\usepackage{lipsum}
\usepackage{braket}
\usepackage{tikz}
\usepackage{pgfplots}
\usepackage{comment}
\usepackage{diagbox}
\usepackage{ifthen}
\usepackage{ragged2e}
\usepackage{multirow}
\usepackage{gensymb}
\usepackage[export]{adjustbox}
\usetikzlibrary{arrows,decorations.pathreplacing,decorations.markings,arrows.meta,patterns,3d}
\usepackage[normalem]{ulem} 
\usepackage{cancel}
\usepackage{microtype}
\usepackage{booktabs}

\theoremstyle{definition}

\usepackage[most]{tcolorbox}

\tcbset {
  base/.style={
    arc=0mm, 
    bottomtitle=0.5mm,
    boxrule=0mm,
    colbacktitle=black!10!white, 
    coltitle=black, 
    fonttitle=\bfseries, 
    left=2.5mm,
    leftrule=1mm,
    right=3.5mm,
    title={#1},
    toptitle=0.75mm, 
  }
}

\definecolor{brandblue}{rgb}{0.34, 0.7, 1}
\newtcolorbox{mainbox}[1]{
  colframe=brandblue, 
  base={#1}
}

\newtcolorbox{subbox}[1]{
  colframe=black!30!white,
  base={#1}
}

\pgfplotsset{compat=1.8}

\begin{document}

\title{Generalized Landau Paradigm for quantum phases and phase transitions}
\date{\today}

\author{Xie Chen}
\affiliation{Department of Physics and Institute for Quantum Information and Matter, \mbox{California Institute of Technology, Pasadena, CA, 91125, USA}}

\begin{abstract} 
The Landau paradigm is a central dogma for understanding phase and phase transitions in condensed matter systems, yet for decades, it has been known that a variety of quantum phases exist beyond the framework. Is there a more general framework that provides a systematic understanding of phases and phase transitions in quantum many-body systems? In this Essay, we discuss how the recently developed notions of generalized symmetry and generalized gauging point to a way to extend the Landau paradigm. In the new framework, {\it beyond-Landau phases and transitions} are attributed to the breaking of generalized symmetries, often induced by the generalized gauging procedure facilitated through the symmetry topological field theory formalism. We discuss what needs to be understood to make the generalized Landau paradigm useful in the study of quantum phase and phase transitions.


\end{abstract}

\maketitle


{\textit{\textbf{The comeback of Landau.---}} In the 1930s, Lev D. Landau established the foundational rules for describing phases and phase transitions in physical systems \cite{Landau1937}. For half a century, this framework was the central dogma in condensed matter physics, covering almost every phase diagram that physicists sought to understand, from water to crystals to superconductors. Cracks started to show in the 1980s with the discovery of the quantum Hall effect whose nontrivial order cannot be attributed to the breaking of any symmetry\cite{Klitzing1980,Tsui1982}. Today, these {\it beyond-Landau} phases and phase transitions are standard topics in research articles and textbooks. Yet, in the last decade, new ideas have emerged that may suggest that everything can still be described as Landau after all \cite{Gaiotto2015, Hofman2019, Delacretaz2020, Iqbal2020, Moradi2022topological,Bhardwaj2023,McGreevy2023}. Can that be true?

Let us wind back and start from the basics. A central question in condensed matter physics is to understand phases and phase transitions. What distinguishes phases? What characterizes the critical point at phase transitions? A particularly fascinating aspect of the answers to these questions is the emergence of \textit{universality}. The universal features of phases and phase transitions are determined by a few simple factors, while no other detail—such as chemical composition, interaction strength, lattice constant—matters. Thanks to universality, knowledge gained from studying phases and phase transitions in one system, even a toy model, can be applied to many other systems within the same so-called {\it universality class}. 

In the original Landau paradigm, {\it symmetry} is the key factor underlying universality. This is best illustrated by the simple example of the phases of water. Water molecules can exist in three phases: gas, liquid, or solid. The gas and liquid phases can be smoothly connected beyond the critical point at $374$\degree C and $218$atm, and are thus considered as the same fluid phase. In contrast, the solid phase is always separated from the fluid phase by a phase transition line. The key difference is that the solid breaks the full translation symmetry by forming a lattice, while the fluid does not. Because of this difference in symmetry,  solids and fluids are sharply distinguishable. Their difference can be quantified by their x-ray diffraction pattern, which is structureless for the fluids but displays peaks for solids. This almost mundane example captures the essential insight of the original Landau paradigm.

\begin{mainbox}{The original Landau paradigm}
For systems whose interactions do not preserve any symmetry,
\begin{itemize}
\item no sharp distinction among different parts of the phase diagram,
\item only one phase exists.
\end{itemize}
For systems whose interactions preserve certain symmetry (forming a group $G$),
\begin{itemize}
\item phases are distinguished by how the symmetry is spontaneously broken,
\item different phases are labeled by the subgroup $H$ of symmetry that is not broken,
\item phases labeled by different $H$ are always separated by phase transitions,
\item a continuous transition can exist between phases with `compatible' symmetries $H_1\subset H_2$,
\item the continuous transition is driven by the fluctuation of an `order parameter' which transforms trivially under $H_1$ but non-trivially under $H_2$.
\end{itemize}
\end{mainbox}

The original Landau paradigm was built upon intuitions about classical systems, such as crystals breaking translation symmetry or ferromagnets breaking spin rotation symmetry. In fact, Pierre Curie, when studying magnetism, had already highlighted the importance of group theory and {\it symmetry breaking} (SB)\cite{Curie1894}. But Landau's formalism also works amazingly well for some intrinsically quantum systems, like superconductors \cite{Landau1965}. On the other hand, the quantum world holds more surprises. The original Landau paradigm is effective when quantum phases are classical-like, in the sense that the state of the system can be qualitatively described without entanglement. When entanglement starts playing a major role in the many-body state, phases and phase transitions exist well beyond what the Landau paradigm predicts. 

Since the discovery of the quantum Hall effect, the study of beyond-Landau phases and phase transitions has become a central topic in quantum condensed matter physics. Universal features of topological \cite{Wen1990,Kitaev2003} and symmetry-protected topological \cite{Gu2009,Pollmann2012} phases have been identified. Continuous transitions between (symmetry-protected) topological phases and incompatible symmetry-broken phases are discovered, where the critical fluctuation often involves emergent fractional degrees of freedom, hence the name {\it deconfined quantum critical point} \cite{Ardonne2004,Senthil2004,Lu2014}. It is generally accepted that, apart from symmetry, topological features in quantum systems, arising from intrinsic many-body entanglement, are crucial in distinguishing phases and characterizing phase transitions. However, many questions are still unanswered: What phases can exist? Which phases can be connected through continuous phase transitions? What are the universal features of these transitions? Much progress has been achieved for free-fermion topological phases, where topology of the band-structure characterizes the phases and transitions occur via band-gap closing. In the presence of strong interactions, however, our understanding remains limited.

\begin{mainbox}{Beyond Landau phases and phase transitions}
For systems whose interactions do not preserve any symmetry,
\begin{itemize}
\item a variety of phases exist, distinguished by their {\it topological order}.
\end{itemize}
For systems whose interactions preserve certain symmetry (forming a group $G$),
\begin{itemize}
\item different phases can exist without breaking any symmetry, distinguished by their {\it Symmetry Protected Topological Order},
\item a continuous transition can exist between symmetry-breaking phases with incompatible symmetries $H_1 \not\subset H_2$. 
\end{itemize}
\end{mainbox}

{\textit{\textbf{The generalized Landau Paradigm.---}} 
While progress in the last 40 years took us away from the original Landau paradigm, recent discoveries suggest that many of these beyond-Landau phases and transitions can still be captured by the Landau paradigm, if we are more flexible about the definition of symmetry in quantum systems and embrace a more general version of the procedure called {\it gauging}.

When we think of symmetries in a quantum many-body setting, the familiar types that come to mind, such as spin rotation, time reversal, and lattice translation, share some common features. They are associated with unitary or anti-unitary operators that act globally on the whole system, with symmetry charges carried by point-like operators. Each of these aspects can fail in the now generalized concept of what constitutes a symmetry\cite{Gaiotto2015,McGreevy2023,Shao2024TASI,SchaferNameki2024}. Generalized symmetries may act only on specific subsystems, leading to subsystem symmetries. The symmetry charge can be carried by extended objects such as lines or membranes, a property of higher-form symmetries. Moreover, generalized symmetry operations may not have an inverse, like unitary or anti-unitary operators do, and are called {\it non-invertible}. Once such generalized symmetries are taken into consideration, some beyond-Landau phases, like certain topological phases, can be interpreted as SB phases as well \cite{Gaiotto2015}.

Gauging is a procedure that promotes a global symmetry in a system to a local symmetry by adding gauge field degrees of freedom and minimally coupling them to the original degrees of freedom. The {\it generalized gauging} procedure can be applied to a generalized symmetry, with more options in choosing the coupling \cite{Fuchs2002,Carqueville2016,Bhardwaj2018,Lootens2023,Lootens2024dualities,Haegeman2015,Kong2020,Gaiotto2021,Chatterjee2023holographic,Huang2023topological,kong2025higher,Bhardwaj2025,seifnashri2025gauging}. While the procedure can be technical, it is most easily explained in the Symmetry Topological Field Theory (SymTFT) / Symmetry Topological Order / Topological Holography formalism \cite{Kong2015,Ji2020,Kong2020,Bhardwaj2020,Pulmann2021,Gaiotto2021,Lichtman2021,kong2020mathematical,Kong2020classification,Apruzzi2023,Chatterjee2023symmetry,Moradi2022topological,Freed2023topological,Lin2023,Kong2018,Kong2021,Kong2022one,Kong2022categories,Kong2024categories,Xu2024,Bhardwaj2025}. The formalism offers a pictorial representation of the underlying categorical structure, making it much easier to explain related ideas.

\begin{figure}[ht]
    \centering
    \includegraphics[scale=0.5]{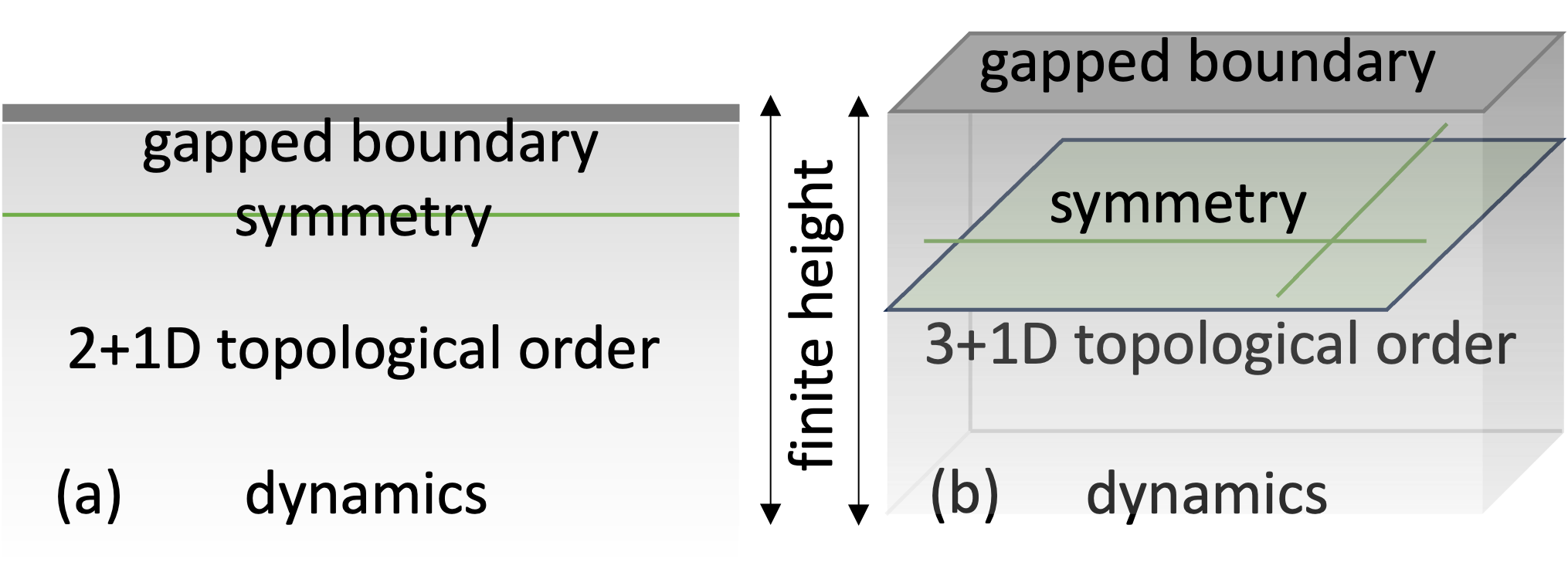}
    \caption{(a) $1+1$D and (b) $2+1$D sandwich structures. A $D+1$-dimensional sandwich has a $(D+1)+1$-dimensional topological order in the bulk, a finite height, and a gapped boundary at the top. The symmetry of the sandwich acts in the bulk, while the bottom boundary contains all the dynamics.
    }
    \label{fig:sandwich}
\end{figure}

In the SymTFT formalism, a $D+1$-dimensional system is realized as a {\it sandwich structure} that has a finite height but is infinite in all other directions (Fig.~\ref{fig:sandwich}). The bulk of the sandwich contains a $(D+1)+1$-dimensional topological order and the top boundary is gapped. They are specifically chosen to realize the symmetry of the $D+1$-dimensional system of interest, while the bottom boundary contains all the (symmetric) dynamics. The sandwich as a whole represents a $D+1$-dimensional system, with its kinematic and dynamic aspects separated vertically. The symmetries of the sandwich emerge as logical operators in the bulk topological state, making it natural to have higher-form symmetries, non-invertible symmetries, etc. The generalized gauging of the symmetry can then be implemented as a change of the top boundary, which leaves the symmetric dynamics unaffected since it is spatially separated from the symmetry \cite{Kong2020,Gaiotto2021,
Chatterjee2023holographic,Huang2023topological,kong2020mathematical,Kong2021}.By combining generalized symmetry and generalized gauging, we arrive at the {\it generalized Landau Paradigm}.

\begin{mainbox}{The generalized Landau paradigm}
\begin{itemize}
\item Topological phases can be interpreted as symmetry-breaking (SB) phases of higher-form symmetries.
\item Topological phase transitions can be described as SB transitions induced by the fluctuation of higher-dimensional charged objects. 
\item Under generalized gauging, symmetry-protected topological phases can be mapped to SB phases of possibly generalized symmetries.
\item The transitions between symmetry-protected topological phases and between incompatible SB phases can be mapped to SB transitions with possibly generalized symmetries.
\end{itemize}
\end{mainbox}
Recently, Lootens {\it et al}.~\cite{Lootens2024} demonstrated, using the tensor network formulation of generalized gauging, that all gapped phases in the sandwich can be mapped to SB phases via generalized gauging, with possibly generalized symmetries. 

In this Essay, I show that all phase transitions in the sandwich can also be mapped to SB transitions, hence significantly expanding the realm of the Landau paradigm. I will first revisit the concept of generalized symmetry, using the $2+1$D $Z_2$ topological phase as an example. Next, I will exemplify the generalized gauging process with the $1+1$D topological superconductor transition. This process is then generalized, for the first time, to encompass all gapped phases and transitions achievable in a sandwich structure. Finally, I will discuss future implications and highlight some of the fascinating open questions that this formalism may help us solve.

{\textit{\textbf{Generalized Symmetry: An example.---}}
\label{sec:1form}
Here, we review how the $2+1$D $Z_2$ topological phase, a phase beyond the original Landau paradigm, can be interpreted as a SB phase of the generalized 1-form symmetry\cite{Gaiotto2015}. The explicit lattice version of the story is given in End Matter (see \textit{Appendix: 1-form symmetry in Toric Code''}), while the discussion here is in continuous space.

\begin{figure}[ht]
    \centering   \includegraphics[scale=0.4]{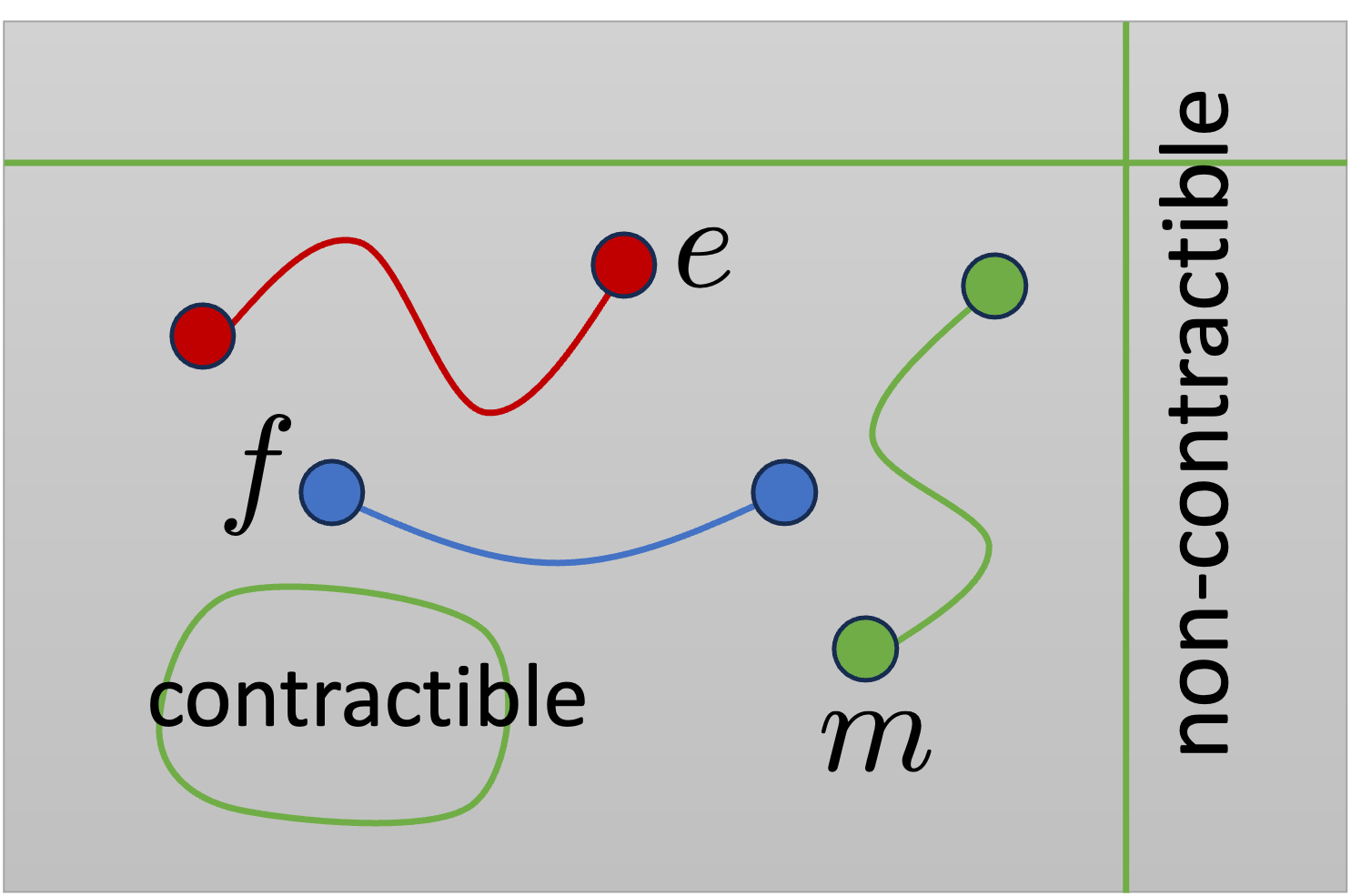}
    \caption{The $2+1$D $Z_2$ topological phase. Fractional excitations $e$, $m$, and $f$ are created at the end of string operators. Closed loop operators, both contractible and non-contractible, generate the $1$-form symmetry of the phase.
    }
    \label{fig:TC}
\end{figure}

The $2+1$D $Z_2$ topological phase contains three types of fractional excitations: $e$ and $m$, which are both bosons, and $f$, which is a fermion. Each type can be created in pairs at the end of a string operator, as shown in Fig.~\ref{fig:TC}. If the string operator forms a closed loop, no excitations are created, and the ground space is mapped back to itself. Therefore, the collection of all loop operators forms a symmetry of the ground space. Each operator acts only on a loop, a submanifold one dimension lower than the full spatial dimension. The symmetry is hence called $1$-form. The conventional global symmetries are referred to as $0$-form. Since two excitations of the same type fuse back to the vacuum, their corresponding $1$-form symmetry is a $Z_2$ symmetry.

If the loop is contractible to a point, its action on the ground space has to be the identity. If the loop goes around a noncontractible cycle in the manifold, it acts nontrivially by mapping between states in the degenerate ground space. This is similar to how the spin rotation symmetry maps between degenerate ferromagnetic ground states in a conventional SB phase. Therefore, the $Z_2$ topological phase is a SB phase of the $1$-form symmetry. 

{\textit{\textbf{Generalized Gauging: An example.---}}
\label{sec:TSc}
Here, we review how the $1+1$D topological superconductor transition\cite{Kitaev2001} can be mapped to a SB transition through the Jordan Wigner transformation, which achieves an equivalent mapping of phases as gauging the fermion parity symmetry\footnote{The Jordan-Wigner transformation differs from gauging by a self-duality of the Ising chain that exchanges the phases on either side of the transition, without changing the nature of the transition.}. The existence of this mapping has been known for nearly a century \cite{Jordan1928}, as I will soon discuss, so this is not new. However, we will present a new approach to identify this mapping using the SymTFT formalism. This approach can be straightforwardly generalized and applied to a variety of other transitions.

{\it How the story is usually told.---}
\label{sec:JW}
Consider a one-dimensional chain of spinless fermions, with two Majorana modes $\gamma_{2j-1}$ and $\gamma_{2j}$ on each site $j$. 
The Hamiltonian for the superconducting chain is:
\begin{equation}
H = -\lambda \sum_{j} i\gamma_{2j-1}\gamma_{2j} - (1-\lambda) \sum_{j} i\gamma_{2j}\gamma_{2j+1}.
\label{eq:HSc}
\end{equation}
The value $\lambda=0.5$ separates the trivial phase from the topological phase, where the critical point corresponds to the massless Majorana chain. A superconducting system always  preserves fermion parity symmetry, $P_f = \prod_j i\gamma_{2j-1}\gamma_{2j}$, which cannot be spontaneously broken. Both the trivial and topological phases preserve this symmetry. Hence, the phase transition is, ostensibly, beyond the Landau paradigm. 

However, it has long been known that this transition can be mapped into a SB one through the so-called {\it Jordan-Wigner transformation}. Under this transformation, the Majorana operators are mapped into (nonlocal) spin $1/2$ operators as
$\gamma_{2j-1} = Z_j\prod_{k<j}X_k, \gamma_{2j} = Y_j\prod_{k<j}X_k$, where $X$,$Y$, $Z$ are $2\times 2$ Pauli matrices.
The fermion parity symmetry operator is mapped to $U = \prod_j X_j$, and the superconducting Hamiltonian is mapped into the Ising chain Hamiltonian\footnote{On a periodic chain, the Ising Hamiltonian differs from the transformed Majorana Hamiltonian by a boundary term. We ignore this subtlety here.}:
\begin{equation}
H = -\lambda \sum_j X_j - (1-\lambda) \sum_j Z_jZ_{j+1}.
\label{eq:HIs}
\end{equation}
$\lambda = 0.5$ separates the symmetric phase from the SB phase of the spin chain model, with the transition described by the Ising critical point. Both the massless Majorana chain and the Ising critical point are known to be described by the two-dimensional Ising conformal field theory (CFT), albeit with minor differences in their partition functions \cite{ginsparg1988applied}. 

A key feature of the SB phase is that when the charge operator $Z$ acts on the degenerate ground space (for example, the space spanned by $|00...0\rangle$ and $|11...1\rangle$ at $\lambda = 0$), it preserves the ground space. In contrast, in the symmetric phase, $Z$ creates a local excitation. This property will be used to identify SB order within the SymTFT formalism.

{\it Gauging facilitated through a bulk.---}
\label{sec:duality}
We now present the SymTFT version of the same mapping. While the SymTFT approach may appear to complicate an otherwise simpler story, it has the advantage of being readily generalizable to other phases and transitions, realizing the generalized gauging procedure. A detailed lattice version of this construction is provided in End Matter (see \textit{Appendix: Lattice realization of the topological superconductor as a sandwich}).

In the SymTFT formulation, the $1+1$D chain is constructed within a sandwich structure with a $2+1$D bulk in the $Z_2$ topological phase. The first step in setting up the sandwich structure is to select a gapped boundary condition at the top and determine the resulting symmetry of the sandwich. There are three types of gapped boundaries in the topological phase, each corresponding to the condensation of one of the three excitations $e$, $m$, and $f$\cite{Bravyi1998arxiv,Lou2021,wen2024topological,Huang2025}. That is, each excitation can be brought to its respective condensed boundary where it disappears into the ground state. The resulting sandwich exhibits a global $Z_2$ symmetry in all three cases, generated by the horizontal loop operators of $m$ (green lines in Fig.~\ref{fig:f2e}(c,d)), $f$, and $e$ (red lines in Fig.~\ref{fig:f2e}(a,b)), respectively. When a fractional excitation tunnels out of its condensed top boundary—say, by the vertical $e$ string operator in Fig.~\ref{fig:f2e} (c)—it flips the value of the symmetry operator, which is the loop operator of a different type. Thus, the vertical string operator acts as a charge operator of the symmetry (analogous to the $Z$ operator in the Ising chain). Due to the finite height of the sandwich, the charge operator is an local operator, and can be used to distinguish between symmetric and SB states.

\begin{figure}[th]
    \centering
    \includegraphics[width=\textwidth]{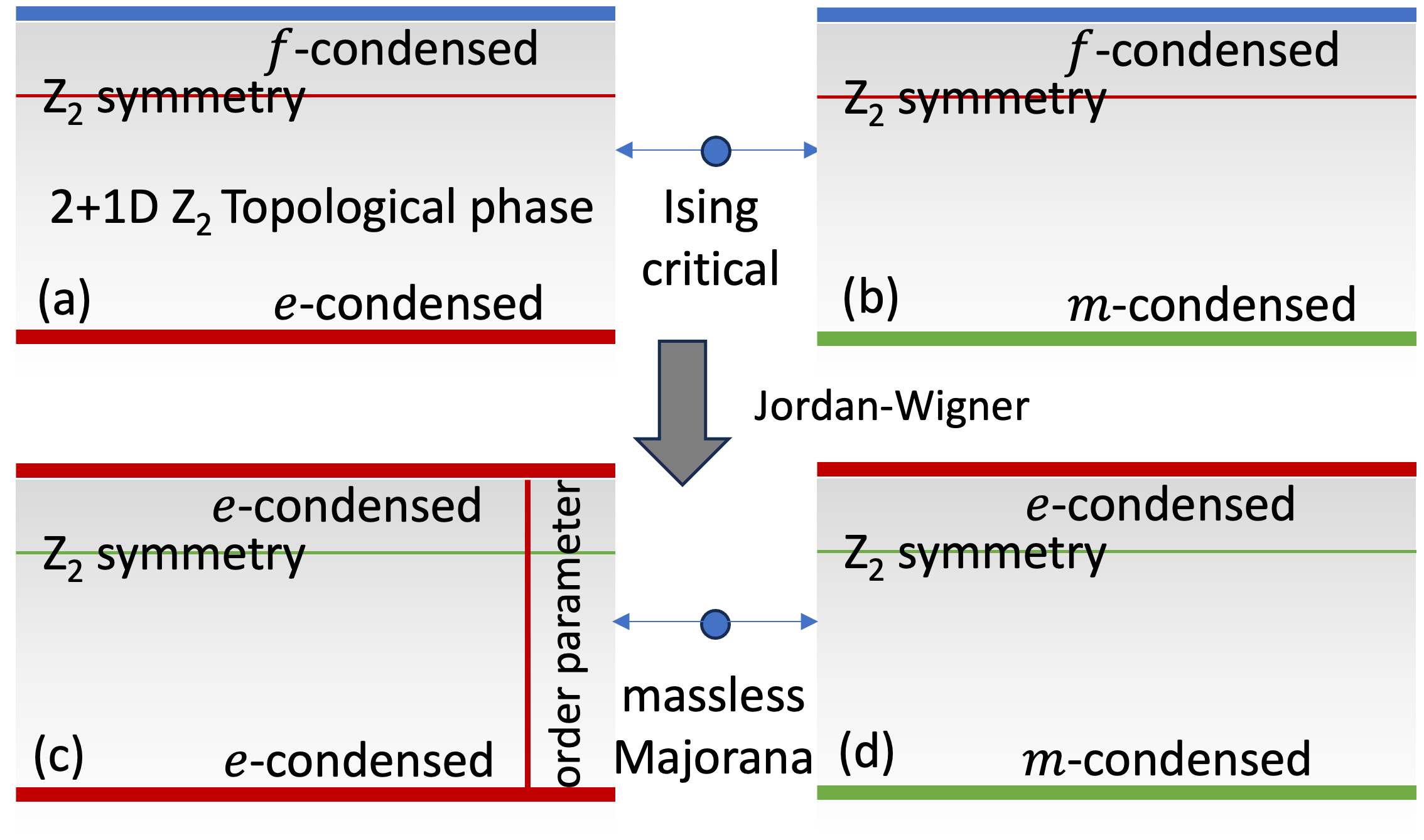}
    \caption{The Jordan-Wigner transformation in the SymTFT formalism. The trivial/topological superconducting chain is realized in a sandwich structure with the $2+1$D $Z_2$ topological order in the bulk, the $f$-condensed boundary at the top, and (a) the $e$-condensed and (b) $m$-condensed boundary at the bottom, respectively. The Jordan-Wigner transformation corresponds to changing the top boundary to $e$-condensed without modifying the bulk or the bottom boundary. The vertical tunneling of the $e$ excitation (red line in c) indicates spontaneous breaking of the $Z_2$ symmetry.}
    \label{fig:f2e}
\end{figure}

To obtain the sandwich description of a fermionic chain, we select the top boundary to be $f$-condensed, which makes the system fermionic (see \textit{Appendix: Lattice realization of the topological superconductor as a sandwich} for details). When the bottom boundary is tuned between $e$-condensed and $m$-condensed states, the charge operator of the fermionic sandwich cannot terminate on the bottom boundary without creating excitations. Consequently, both $e$- and $m$-condensed bottom boundaries correspond to symmetric phases under the $Z_2$ fermion parity symmetry. This is hardly surprising since we know there are two symmetric phases under the $Z_2$ fermion parity symmetry,  the trivial phase and the topological superconductor. Tuning between the $e$-condensed and $m$-condensed boundaries at the bottom leads to the massless Majorana chain at the transition point.

To map the fermionic system to a spin chain, all we need to do is change the top boundary from $f$-condensed to $e$-condensed, as shown in Fig.~\ref{fig:f2e}. The topological (trivial) superconductors are mapped to the SB (symmetric) phases with a spin $Z_2$ symmetry, where the charge operator can (cannot) end at the bottom boundary without producing excitations. The massless Majorana chain is mapped to the critical Ising chain. The difference between the partition functions of these two critical points can be directly observed from the sandwich construction. The Ising CFT sector involved in the partition function is determined by the top boundary: For the $f$-condensed top boundary, the partition function is a sum over the vacuum and the fermion sectors; for the $e$-condensed top boundary, the partition function is a sum over the vacuum and the spin sectors. 

{\it{\textbf{Unified picture through SymTFT.---}}}
\label{sec:SymTFT}
The sandwich structure and the {\it change-of-top-boundary procedure} described above can be applied to a wide range of phases and phase transitions. This approach elucidates the generalized symmetry in the system and facilitates generalized gauging, enabling beyond-Landau phases and phase transitions to be mapped back into the SB paradigm.

{\it The sandwich and generalized symmetries.---}
\label{sec:SymTFT_symmetry}
First, let's discuss which types of generalized symmetries can be realized in the sandwich structure. We focus on $1+1$D and $2+1$D sandwich structures with $2+1$D and $3+1$D topological orders in the bulk (as shown in Fig.~\ref{fig:sandwich}). Comprehensive discussions of these two cases can be found in Ref.~\cite{Bhardwaj2025,bhardwaj2025lattice} and Ref.~\cite{bhardwaj2024gapped1,bhardwaj2025gapped2,wen2025topological,inamura202521d}. 

By now, much is known about topological orders in $2+1$D. First, note that $2+1$D topological orders in the bulk of a $1+1$D sandwich cannot be chiral, because chiral states do not possess gapped boundaries. Among nonchiral theories, let's consider $2+1$D gauge theories with a discrete gauge group $G$. Bulk fractional excitations include gauge charges and gauge fluxes. If all gauge charges condense on the top boundary, the string operators of the gauge charges parallel to the boundary are forced to take specific values while the string operators of the gauge fluxes give rise to the symmetry of the sandwich, a regular $0$-form group $G$ symmetry. If, instead, the gauge flux excitations condense on the top boundary, the symmetry is generated by the string operators of the gauge charges and forms a Rep($G$) symmetry. That is, the symmetry operators are labeled by the irreducible representations of $G$ and satisfy the same fusion (multiplication) rule. When $G$ is non-Abelian, the Rep($G$) symmetry is non-invertible. For example, when $G = S_3$, Rep($S_3$) contains three elements—$W_0$, $W_1$, and $W_2$—corresponding to the trivial, sign, and 2-dimensional irreps of $S_3$, respectively, which satisfy the fusion rule:

\begin{equation}
W_2\times W_2 = W_0 + W_1 + W_2.
\end{equation}
Hence, $W_2$ is a non-invertible symmetry operator. 

Not all $2+1$D topological orders are gauge theories of finite groups. With other types of non-abelian topological orders in the bulk, it is possible to have other interesting non-invertible symmetries in the sandwich. For example, with the doubled-Fibonacci topological order in the bulk, we can have the Fibonacci symmetry, $W_{\tau}$, satisfying:
\begin{equation}
W_{\tau}\times W_{\tau} = W_I + W_{\tau},
\end{equation}
where $W_I$ is the identity operator. The phase diagram with the Fibonacci symmetry has been studied as a prominent example of the {\it Golden Chain} \cite{Feiguin2007}.

The situation is, in fact, simpler for $2+1$D sandwiches, because all gapped topological orders in $3+1$D are gauge theories with discrete gauge groups\cite{Johnson-Freyd2022, Lan2018, Lan2019}. In this case, the gauge charge excitations remain point-like, while the gauge flux excitations are loop-like and are created at the boundaries of membrane operators. If all the gauge charges condense on the top boundary, the membrane operators parallel to this boundary become the $0$-form symmetry operators of the sandwich, forming the group $G$. If, instead, the gauge flux excitations condense on the top boundary, the symmetry is generated by the string operators of the gauge charges in the bulk, forming a Rep($G$) $1$-form symmetry.  For example, when $G=Z_2$, the sandwich exhibits a $Z_2$ 1-form symmetry. If the bottom boundary also condenses gauge fluxes, this 1-form symmetry is spontaneously broken, and the sandwich corresponds to a $2+1$D $Z_2$ gauge theory. Finally, if the top boundary condenses a mixture of gauge charge and gauge flux, the sandwich has both $0$-form and 1-form symmetries. 

Are there conventional symmetries and gapped phases that do not fit into a sandwich structure? In $1+1$D, all gapped phases with finite internal (conventional) symmetries can be realized in a sandwich structure. In $2+1$D, certain non-gauge theory topological orders, such as the doubled Fibonacci, cannot be realized in a sandwich, at least not with minimal boundaries where all fractional excitations are connected to fractional excitations in the bulk. Here, we focus on sandwiches with a gapped topological bulk, so that their associated generalized symmetries are finite internal symmetries. There are interesting proposals extending the SymTFT formalism to continuous as well as spatial symmetries. We refer the readers to Ref.~\cite{apruzzi2024symth,Antinucci2025,brennan2024symtft,bonetti2024symtfts,pace2025space} for more discussion.

{\it The sandwich and generalized gauging.---} 
\label{sec:SymTFT_gauging}
The gauging procedure illustrated previously—where the top boundary condition in a sandwich is changed—can now be applied more generally, introducing the notion of {\it generalized gauging}. 

\begin{figure}[th]
    \centering
    \includegraphics[width=\textwidth]{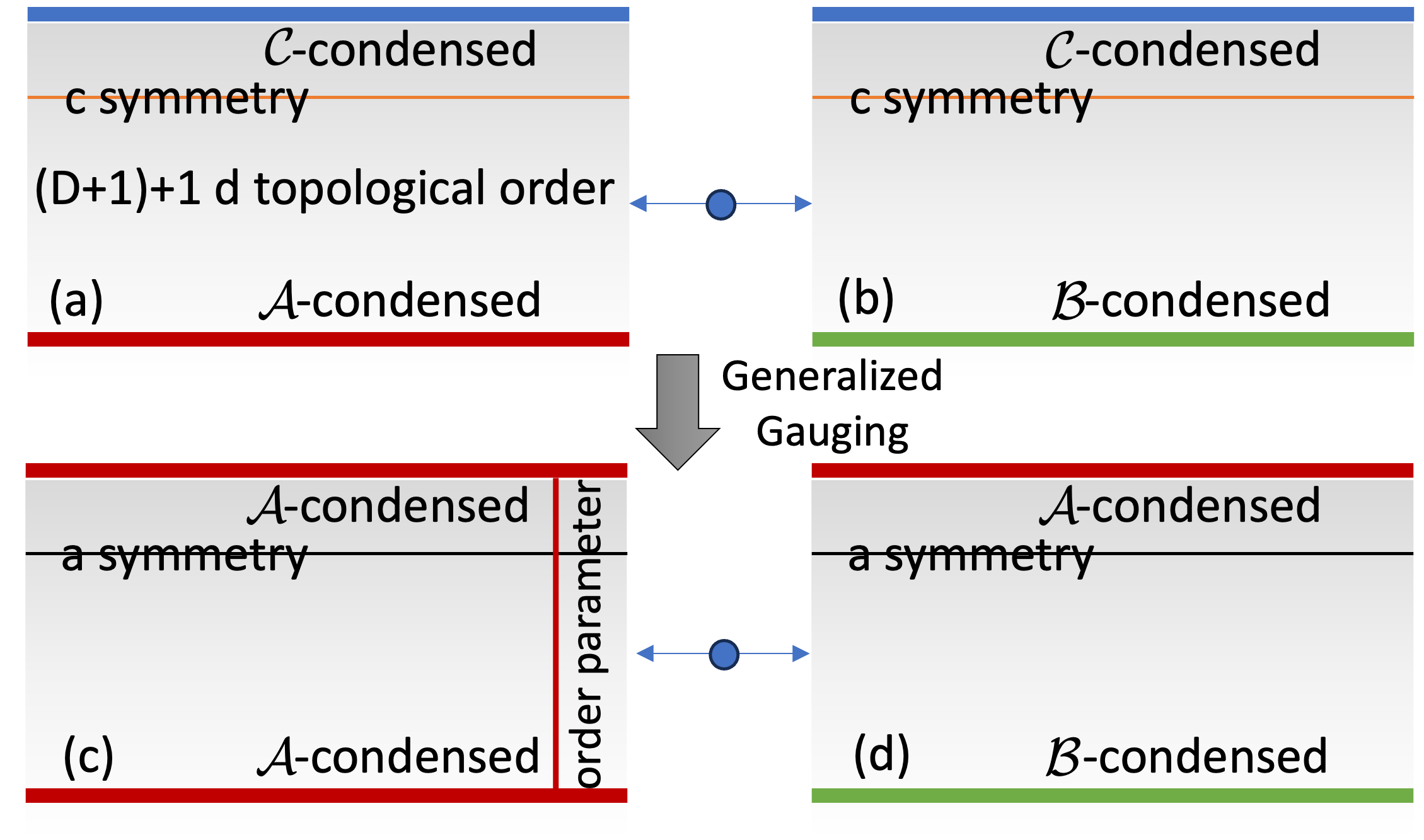}
    \caption{Generalized gauging (changing the top boundary from $\mathcal{C}$-condensed to $\mathcal{A}$-condensed) maps general gapped phases (a) to SB phases (c) and the phase transition between general gapped phases (a,b) to SB transitions (c,d). Blue dots indicate transition points between gapped phases. The red line represents the tunneling of fractional excitations between matching condensates on the two boundaries, which becomes the order parameter in the spontaneous breaking of $a$-symmetry.}
    \label{fig:C2A}
\end{figure}

Consider a $D+1$-dimensional gapped phase realized in a sandwich with a $(D+1)+1$-dimensional topological order in the bulk, a $\mathcal{C}$-condensed gapped boundary at the top, and an $\mathcal{A}$-condensed gapped boundary at the bottom (Fig.~\ref{fig:C2A} (a)). Such a phase can be mapped to a fully SB phase by applying the generalized gauging procedure where the top boundary is changed from $\mathcal{C}$-condensed to $\mathcal{A}$-condensed, effectively shifting the symmetry from $c$-symmetry to $a$-symmetry (as illustrated from Fig.~\ref{fig:C2A} (a) to (c)). 
After this mapping, all fractional excitations condensed in $\mathcal{A}$ can tunnel from the top to the bottom boundary without creating any excitation. These vertical operators are charged under the $a$-symmetry and become the order parameter of the $a$-SB phase. In this way, all gapped phases (that can fit into a sandwich) can be gauged into fully SB phases, with potentially generalized symmetries\cite{Lootens2024}. 

Now, if we consider a second gapped phase realized in the same sandwich—that is, with the same bulk and the same top boundary—but with a $\mathcal{B}$-condensed bottom boundary (see Fig.~\ref{fig:C2A} (b)), the transition between these two phases is mapped to a SB transition when the top boundary is changed from $\mathcal{C}$-condensed to $\mathcal{A}$-condensed. After this generalized gauging procedure, the phase in (c) is fully SB, while the phase in (d) is not because fewer excitations can tunnel between the top and bottom boundaries (see exceptions explained in End Matter, \textit{Appendix: Exceptions to the mapping in Fig.~\ref{fig:C2A}}). Therefore, the transition from (c) to (d) can be induced by the fluctuation of the order parameter of the generalized $a$-symmetry, a phenomenon that could be described by a properly generalized Landau theory. 

Beyond the  Jordan-Wigner, known examples that fit this framework include the Kennedy-Tasaki transformation \cite{Kennedy1992a,Kennedy1992b}, which maps the transition between the $Z_2\times Z_2$ SPT phases to the $Z_2\times Z_2$ SB transition\cite{Huang2023topological}, as well as the transformation mapping the deconfined critical point between partial SB phases of the anomalous $Z_2\times Z_2$ symmetry to the SB transition of $Z_4$ symmetry\cite{Zhang2023}. Many new cases await to be explored.
 
{\textit{\textbf{Where do we go from here?---}}
\label{sec:outlook}
We have shown how, by introducing generalized symmetries and generalized gauging, the Landau paradigm is generalized. Beyond-Landau phases become SB phases of potentially generalized symmetries, and beyond-Landau phase transitions become SB transitions. This provides a more systematic framework for understanding phases and phase transitions. What open questions can be addressed using this framework? What new questions arise from this generalized perspective? Below, we discuss a few possibilities.


{\it Phases with generalized symmetries.---}
It is interesting to explore what kinds of phases can exist in systems with generalized symmetries. 

1. Is there always a gapped symmetric phase? The answer is known to be no for certain symmetries because of underlying anomalies \cite{Bhardwaj2018,Chang2019,Thorngren2019Fusion,Choi2022,Choi2023,kaidi2023symmetry,Cordova2024,Copetti2024}. What kinds of anomalies can generalized symmetries possess? What are the general criteria for detecting such anomalies? 

2. What are the possible symmetry-breaking phases\cite{etxebarria2022goldstone,Damia2024,Bhardwaj2025}? For conventional symmetries, phases are labeled by the unbroken subgroup of the symmetry. For generalized symmetries, what are the possibilities of unbroken symmetries? Moreover, there can sometimes be more than one SB phase even when the symmetry is fully broken. For example, in the case of a 1-form $Z_2$ symmetry in $2+1$D, two distinct $Z_2$ gauge theories both spontaneously break the 1-form symmetry, but they are different phases. What other quantities are needed to distinguish these SB phases? 

3. Are there nontrivial symmetry-protected topological phases (SPT) under the generalized symmetry? For instance, nontrivial SPT phases of 1-form symmetries exist in $3+1$D\cite{Tsui2020}. With non-invertible symmetries in $1+1$D systems, multiple symmetric phases can exist, but their structure differs drastically from that of invertible symmetries, as the former do not necessarily form an abelian group \cite{Seifnashri2024,Fechisin2025}.}

The SymTFT formalism provides a convenient tool to address these questions. Once the bulk and the top boundary are chosen for a sandwich structure to realize certain generalized symmetries, the classification of phases translates to classifying the possible bottom boundary conditions, which can be studied independently of the top boundary. 


{\it Symmetry-breaking transition of generalized symmetries.---}
For phase transitions, the implications of the generalized Landau paradigm can be both conceptual and practical.
On the conceptual level, it allows us to more systematically address the question: Which phases can be connected through a continuous phase transition? Continuous transitions occur when a physical mechanism drives the deformation of the system from one phase to another. In the original Landau paradigm, this mechanism is the fluctuation of the order parameter, and it works between SB phases with compatible symmetries. Over the past decades, studies of phases and transitions beyond Landau’s framework have revealed many new mechanisms that can link seemingly incompatible phases, though these mechanisms are usually case-specific. The generalized Landau paradigm, however, has the potential to identify a unifying mechanism for continuous phase transitions between phases that can fit into a sandwich structure, as the fluctuation of the order parameter associated with generalized symmetries.

Of course, many practical questions remain open. Is the transition inherently continuous, or does it require extra symmetry to ensure its continuousness? What type of field theory describes the transition? What are the universal critical properties across the transition? Given that such transitions are driven by the fluctuations of the order parameter, one might attempt to follow Landau's procedure and develop a field theory description of the transition based on an {\it order parameter field}. This approach, together with the renormalization group method developed by Wilson later on \cite{Wilson1971a,Wilson1971b}, has proved highly successful within the original Landau paradigm. Can this success be extended to the generalized Landau paradigm? This is an interesting open direction. 

Some progress has been made in formulating a field theory for the 1-form SB transition from the trivial to the topological phase \cite{Iqbal2020,Iqbal2022}, though analysis is challenging since the order parameter is an extended object. A potentially simpler still interesting case is the SB of 0-form non-invertible symmetries in 1+1D systems, where the order parameters remain point-like. Would it be possible to develop a field theory description based on these order parameters to identify universality classes beyond the original Landau paradigm? This remains an intriguing and open question.

{\it Longer term prospect of SymTFT to study phase and phase transitions.---}
The SymTFT formalism plays an important role in the discussion of this Essay. It makes the appearance of generalized symmetries more natural and the implementation of generalized gauging more straightforward. Its strength comes from spatially separating the system’s kinematic part (the symmetry) from the dynamic part (Hamiltonian terms) through the sandwich structure. This structure can provide insights into critical theories that may not be obvious otherwise. 

In recent decades, substantial progress has been made in understanding two main classes of quantum systems: The $1+1$D critical systems via the 2d CFT, and the $2+1$D gapped systems through topological quantum field theory. The SymTFT formalism bridges these approaches and clarifies how to extract the kinematic aspects of the $1+1$D critical theory into a $2+1$D topological bulk. The same approach may prove to be useful to shed light on the structure of critical points in $2+1$D and higher, where a CFT understanding is much more limited. In particular, the topological bulk helps relate the emergent symmetries in gapped phases and those at their transition points. In $1+1$D, these emergent symmetries can manifest as primary fields in the CFT, whereas in $2+1$D and higher, due to the lack of state-operator correspondence for extended operators, emergent symmetries may only be evident in the energy spectrum \cite{ji2024topological} rather than the correlation functions \cite{Zhao2021}.


\begin{acknowledgments}
Xie Chen is grateful for helpful discussions with Nathanan Tantivasadakarn, Sakura Schafer-Nameki, John McGreevy, Frank Verstraete, Xiao-Gang Wen, Nathan Seiberg, Liang Kong, and Shu-Heng Shao. X.C. is supported by the Simons collaboration on `Ultra-Quantum Matter'' (grant number 651438), the Simons Investigator Award (award ID 828078), the Institute for Quantum Information and Matter at Caltech, and the Walter Burke Institute for Theoretical Physics at Caltech.
\end{acknowledgments}

\bibliography{references}

\appendix

{\textit{\textbf{Appendix: 1-form symmetry in Toric Code.---}}
\label{sec:TC}
Consider the Toric Code model\cite{Kitaev2003} on a two-dimensional square lattice with periodic boundary conditions (Fig.\ref{fig:TCl}). Each edge in the lattice hosts a spin $1/2$ degree of freedom. 

\begin{figure}[th]
    \centering
    \includegraphics[scale=0.5]{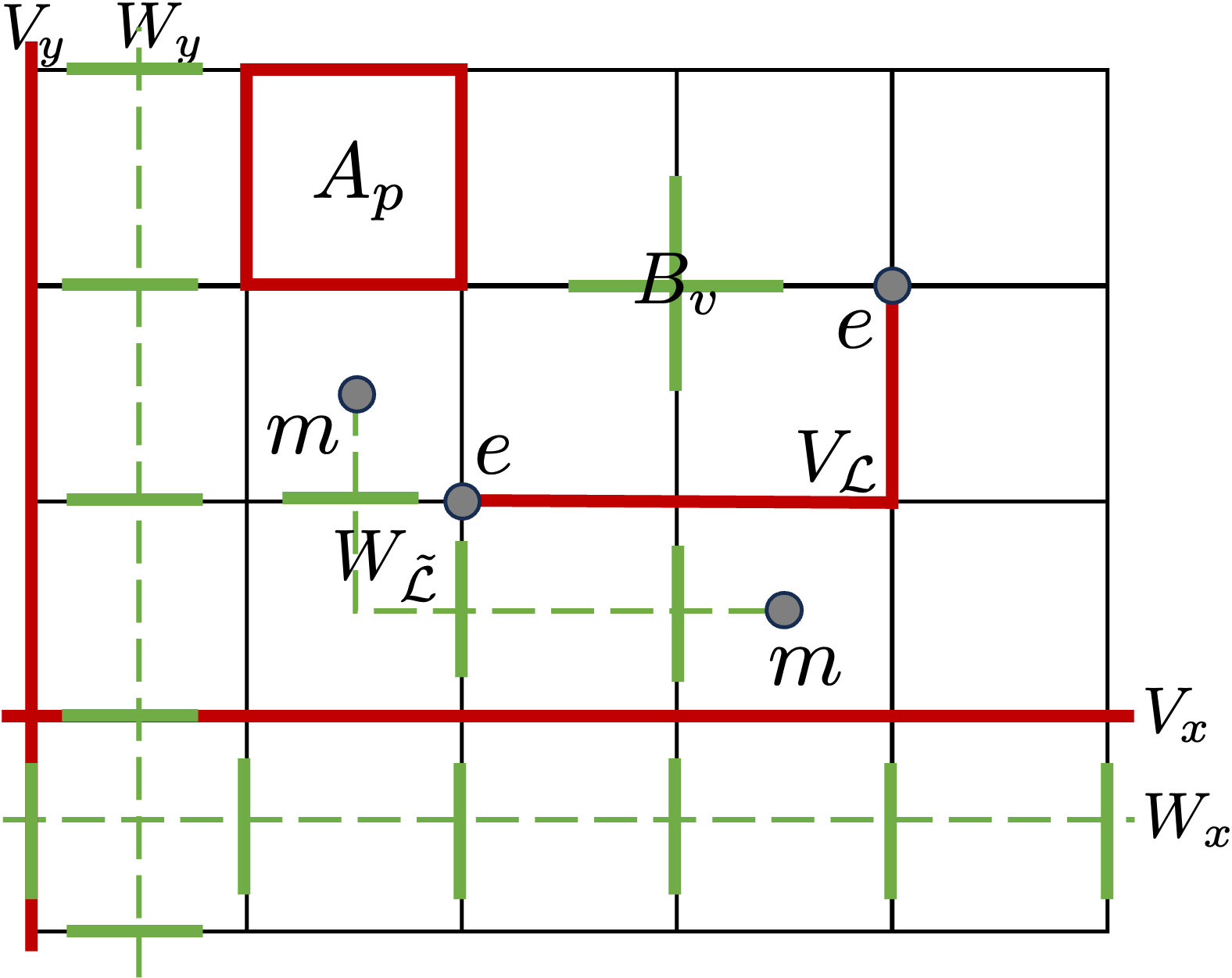}
    \caption{The Toric Code model on a two-dimensional square lattice. Red edges represent the $X$ operator, green edges represent the $Z$ operator. The Hamiltonian is a sum of the plaquette terms $A_p$ and the vertex terms $B_v$. Closed string operators along nontrivial cycles $W_x$,$W_y$,$V_x$,$V_y$ act non-trivially in the ground space. Open string operators $V_{\mathcal{L}}$ and $W_{\tilde{\mathcal{L}}}$ create fractional excitations $e$ and $m$ at the two ends. }
    \label{fig:TCl}
\end{figure}

The Hamiltonian is a sum of plaquette terms $A_p$ and vertex terms $B_v$. $A_p$ is a product of $X$ operators on the edges around the plaquette $p$ and $B_v$ is a product of $Z$ operators on the edges ending at vertex $v$.
\begin{equation}
\begin{array} {lll}
H & = & -\sum_p A_p - \sum_v B_v - h \sum_e Z_e \\
& = &
-\sum_p \prod_{e\in p} X_e - \sum_v \prod_{v \in e} Z_e - h \sum_e Z_e
\end{array}
\label{eq:HTC}
\end{equation}
We have added a magnetic field term with magnitude $h$. When $h=0$, the ground state satisfying $A_p=1$, $B_v=1$ is gapped and has $Z_2$ topological order. The $Z_2$ topological order is characterized by the fractional charge excitation $e$ ($B_v = -1$) and the fractional flux excitation $m$ ($A_p = -1$) generated at the ends of open string operators $V_{\mathcal{L}} = \prod_{e \in \mathcal{L}} X_e$ and $W_{\tilde{\mathcal{L}}} = \prod_{e \in \tilde{\mathcal{L}}} Z_e$. $\mathcal{L}$ and $\tilde{\mathcal{L}}$ are open strings on the lattice and dual lattice, respectively, represented by the red and green open strings in Fig.~\ref{fig:TCl}. $e$ and $m$ are both bosonic excitations, while their composite $f$ is fermionic.

The $Z_2$ topological order does not require any symmetry protection. That is, the system remains in the same phase in the presence of any local perturbation as long as the perturbation is weak enough. Nonetheless, the particular Hamiltonian in Eq.~\ref{eq:HTC} does have many symmetries (even with nonzero $h$). In particular, it is invariant under all loop operators of the form
\begin{equation}
W_{\tilde{\mathcal{C}}} = \prod_{e\in \tilde{\mathcal{C}}} Z_e,
\label{eq:Wc}
\end{equation}
where $\tilde{\mathcal{C}}$ is any closed loop on the dual lattice. The $B_v$ terms are the smallest loop operators in this set. On the torus, there are also nontrivial loop operators $W_x$ and $W_y$ represented by the horizontal and vertical green lines in Fig.~\ref{fig:TCl}. The full set of operators in Eq.~\ref{eq:Wc} form the 1-form symmetry of the model. 1-form means that the operators each act on a sub-manifold (a loop) that is one dimension lower than the full spatial dimension of the system. The conventional global symmetries are referred to as 0-form symmetries in this terminology.

Note that the 1-form symmetry should be treated as a {\it topological symmetry}, meaning that operators that can be smoothly deformed into each other in the continuum limit should act in the same way. In particular, loop operators around contractible loops should all act trivially $W_{\text{contractible } \tilde{\mathcal{C}}} = 1$. Only $W_x$ and $W_y$ can potentially act nontrivially. We are hence restricted to consider a highly constrained Hilbert space. Operators charged under the 1-form symmetry are the dual loop operators $V_x$ and $V_y$, which are products of $X$ along the nontrivial cycles on the direct lattice (shown as red lines in Fig.~\ref{fig:TCl}). $V_x$ carries nontrivial charge under $W_y$ while $V_y$ carries nontrivial charge under $W_x$. See Ref.~\cite{Oh2023,Choi2025} for more discussion about topological vs. non-topological 1-form symmetries. 

When $h=0$, the model has a four-fold degenerate ground space and the $W_x$ and $W_y$ operators act non-trivially in the space. In particular, if we choose a basis for the ground space to be the eigenstates of $V_x$ and $V_y$, the basis states map into each other under $W_x$ and $W_y$ and break the 1-form symmetry. Therefore, the Toric Code, and its associated $Z_2$ topological order, can be interpreted as an SB phase of the 1-form symmetry. $W_x$, $W_y$, $V_x$, and $V_y$ are the `logical' operators that act in the degenerate ground space. When $h>>1$, all spins are in the $|0\rangle$ state. The topological order is lost, and the system is in the trivial `Higgs' phase. With respect to the 1-form symmetry, 
the unique ground state is invariant under all 1-form symmetry operators. Therefore, as $h$ increases from zero, the model transitions from a SB phase of the 1-form symmetry to a symmetric phase. 

This feature of higher-form SB, illustrated with the Toric Code model, applies to a wide class of topological phases. Another interesting example of 1-form SB is the $3+1$D quantum electrodynamics theory. In the absence of charge excitations, the system has a $U(1)$ 1-form symmetry. The gapless photon excitation in the Coulomb phase of the theory can be interpreted as the Goldstone mode resulting from the spontaneous breaking of the continuous $U(1)$ 1-form symmetry. We refer the readers to Ref.~\cite{Gaiotto2015} for a detailed explanation of this example. In this essay, we focus mostly on finite symmetries. Some earlier work on interpreting topological order and the gaplessness of photons using generalized symmetries can be found in Ref.~\cite{Nussinov2009a,Nussinov2009b,Kovner1994}

{\textit{\textbf{Appendix: Lattice realization of the topological superconductor as a sandwich.---}}
\label{sec:f2e_lattice}
As shown in Fig.~\ref{fig:TCb}, there are three types of gapped boundary conditions in the Toric Code, corresponding to the condensation of the three fractional excitations $e$, $m$, and $f$\cite{Bravyi1998arxiv,Lou2021,wen2024topological,Huang2025}. The sandwich structure has a finite height in the $y$ direction and an infinite length in the $x$ direction, and we will assume periodic boundary conditions in the $x$ direction (the figure shows only a patch near the top boundary). The figure shows (part of) the bulk, which is in the ground state of the Toric code Hamiltonian shown in Fig.~\ref{fig:TCl}, and the top boundary. 

\begin{figure*}[th]
    \centering
    \includegraphics[width=0.9\textwidth]{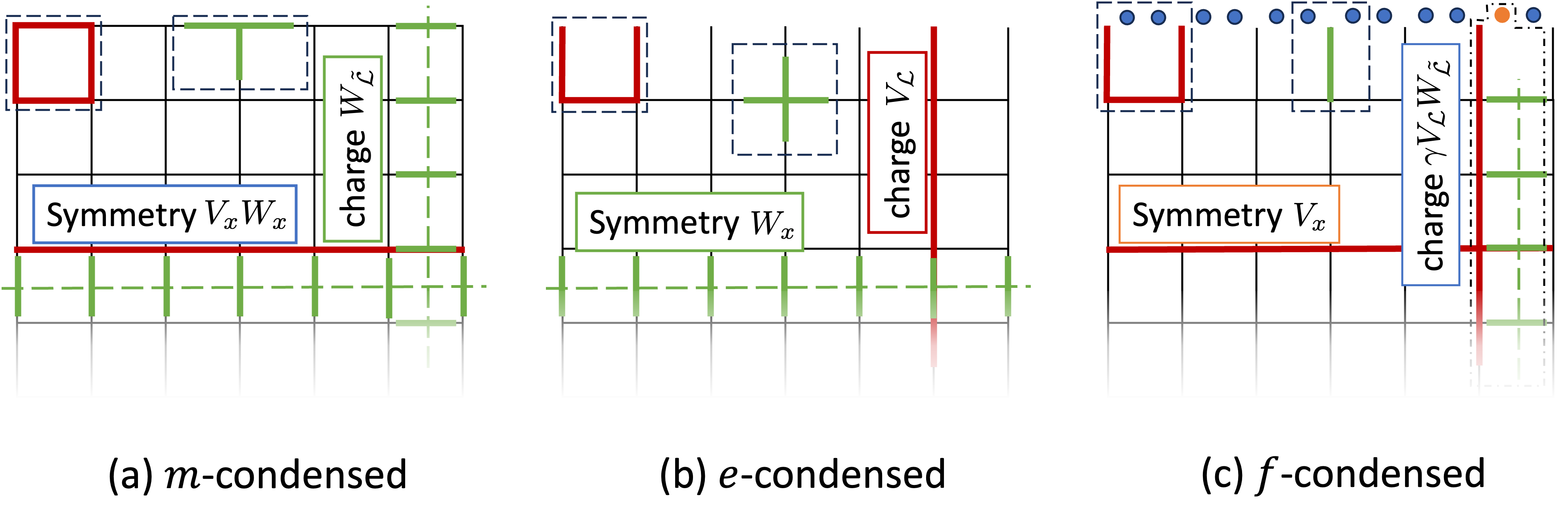}
    \caption{Three types of gapped boundary of the Toric Code model. Red edges represent the $X$ operator, green edges represent the $Z$ operator, dots represent Majorana modes $\gamma$. Hamiltonian terms at the boundary are given in dashed boxes. The horizontal (a) $V_x$, (b) $W_x$, (c) $V_x$ operators generate the symmetry of the resulting sandwich. The vertical open string operators (a) $W_{\tilde{\mathcal{L}}}$, (b) $V_{\mathcal{L}}$, (c)$\gamma V_{\mathcal{L}}W_{\tilde{\mathcal{L}}}$ (enclosed in the dash-dotted box) are the charge operators under the respective symmetry.}
    \label{fig:TCb}
\end{figure*}

Fig.~\ref{fig:TCb} (a) represents the case where $m$ condenses. The Hamiltonian terms at the top boundary are given in the dashed boxes and include the red four-body terms of $\prod_e X_e$ around a plaquette as well as the green three-body terms of $\prod_e Z_e$ around a vertex. $m$ is said to condense on the boundary because a $W_{\tilde{\mathcal{L}}}$ string operator, which brings an $m$ excitation from the bulk to the top boundary, can terminate on the boundary without creating any excitation. This string operator becomes the symmetry charge operator of the sandwich, as explained below.

The product of all the three-body $\prod_e Z_e$ terms along the top boundary gives $W_x$, the string operator of $m$ along the $x$-direction nontrivial cycle. Therefore, as long as no excitation exists at the top boundary (or in the bulk), $W_x = 1$. $V_x$ (the red horizontal line in Fig.~\ref{fig:TCb} (a)), on the other hand, is not constrained in this sandwich Hilbert space. It commutes with all the Hamiltonian terms in the sandwich, including any term at the bottom boundary not shown in this figure, simply because they are spatially separated from $V_x$. Therefore, $V_x$ generates a $Z_2$ symmetry of the effective $1+1$D system. The $W_{\tilde{\mathcal{L}}}$ operator intersects with $V_x$ and anti-commutes with it. Therefore, $W_{\tilde{\mathcal{L}}}$ becomes the charge operator under the $Z_2$ symmetry. Note that, since the sandwich has a finite height, $W_{\tilde{\mathcal{L}}}$ is a local (point-like) charge operator.

Similar analysis applies to Fig.~\ref{fig:TCb} (b) and (c). In Fig.~\ref{fig:TCb} (b), the Hamiltonian terms at the top boundary include the red three-body $\prod_e X_e$ terms around an open plaquette and the green four-body $\prod_e Z_e$ terms around a vertex. A vertical $V_{\mathcal{L}}$ operator can end on the top boundary without creating any excitation. Therefore, the top boundary is a condensate of $e$. A $Z_2$ symmetry of the sandwich is generated by the $W_x$ operator (dashed green horizontal line) while the $V_{\mathcal{L}}$ operator is the charge operator. In Fig.~\ref{fig:TCb} (c), in order to generate an $f$ condensed boundary, we need to add two Majorana modes (dots) per unit cell at the top boundary. The Hamiltonian terms around an open plaquette are a product of three $X_e$'s together with the two Majorana operators above the plaquette ($i\gamma_{2j-1}\gamma_{2j}$). Another term is the product of $Z_e$ on a boundary vertical edge together with the two Majorana operators above the edge ($i\gamma_{2j}\gamma_{2j+1}$). The composite string operator $V_{\mathcal{L}}W_{\tilde{\mathcal{L}}}$ can terminate at the top boundary without creating any excitation if it is dressed by a Majorana operator $\gamma$ (the orange dot) at the end. The $f$ excitation (or, strictly speaking, the composite of $f$ and a Majorana fermion) is therefore condensed at the top boundary. An $Z_2$ symmetry is generated by the $V_x$ operator (the red horizontal line) while $\gamma V_{\mathcal{L}}W_{\tilde{\mathcal{L}}}$ (enclosed in the dash-dotted box) is the charge operator. The symmetry is hence the fermion parity symmetry of the fermionic chain. 

The second step in setting up the sandwich is to decide what happens at the bottom boundary. Once the bottom boundary is chosen, we have the full description of the sandwich and will be able to know what effective $1+1$D theory it is describing. Since the bottom boundary is spatially separated from the bulk and the top boundary, which determines the symmetry, any Hamiltonian term chosen for the bottom boundary will automatically preserve the symmetry. If the bottom boundary is also chosen to be gapped, the sandwich describes a one-dimensional gapped phase. For example, if the top boundary is $e$-condensed and the bottom boundary is also $e$-condensed, we have a gapped system with $Z_2$ symmetry. Is it in the symmetric phase or the SB phase? Since the charge operator $V_{\mathcal{L}}$ can terminate at both the top and bottom boundaries without creating any excitation, this is the SB phase. On the other hand, if the bottom boundary is $m$-condensed, the system is in the symmetric phase. If the bottom boundary is tuned between the $e$-condensed and the $m$-condensed boundaries, the Ising critical point is encountered in between.

{\textit{\textbf{Appendix: Exceptions to the mapping in Fig.~\ref{fig:C2A}.---}}
\label{sec:exception}
The simplest known example in $1+1$D involves a group $G$ of order 128\cite{Pollmann2012a,Davydov2014,kobayashi2025soft}. This symmetry can be realized in a sandwich structure with the $2+1$D topological order of the quantum double of $G$ in the bulk. This topological order has two types of gapped boundaries $\mathcal{A}$ and $\mathcal{B}$ which condense the same set of anyons but differ by certain phase factors at their fusion vertices. If we start with two sandwiches with $\mathcal{A}$ and $\mathcal{B}$ at the bottom boundary, respectively, applying the generalized gauging procedure illustrated in Fig.~\ref{fig:C2A} results in two sandwiches where the same set of anyons can tunnel between the top and bottom boundaries. Therefore, both correspond to fully SB phases. 

In dimensions higher than $1+1$D, it is easier to find examples where the same set of bulk excitations condense in different ways on the boundary, resulting in different gapped boundary conditions. For example, the $3+1$D $Z_2$ topological order has a smooth and a twisted smooth boundary, where the flux loop condenses but in different ways\cite{Zhao2023,Luo2023,Ji2023}. Starting with two sandwiches with these two types of boundaries at the bottom, the procedure in Fig.~\ref{fig:C2A} maps to two systems both with spontaneously broken $Z_2$ 1-form symmetry. To what extent the transitions between such pairs of systems can be interpreted as SB transitions is not entirely clear and remains to be explored. 

\end{document}